\documentclass[12pt]{iopart}
\usepackage[utf8x]{inputenc}
\usepackage{graphicx}
\usepackage{hyperref}
\usepackage{cleveref}
\usepackage{bm}
\usepackage{color}

\usepackage{iopams}
\begin{document}

\title{Properties of Hybrid Stars with Density Dependent Bag Model}

\author{Debashree Sen$^1$, Naosad Alam$^1$, and Gargi Chaudhuri$^{1,2}$}

\address{$^1$Physics Group, Variable Energy Cyclotron Centre, 1/AF Bidhan Nagar, Kolkata 700064, India}
\address{$^2$Homi Bhabha National Institute, Training School Complex, Anushakti Nagar, Mumbai 400085, India}
\ead{debashreesen88@gmail.com, naosadphy@gmail.com, gargi@vecc.gov.in}
\vspace{10pt}
\date{\today}

\begin{abstract}

 The phenomena of deconfinement of hadronic matter into quark matter at high density, relevant to hybrid star (HS) cores, is studied in the present work. The effective chiral model describes the pure hadronic phase while for the quark phase the MIT bag model is chosen with density dependent bag pressure. Phase transition is achieved using Maxwell construction. The effect of variation of the asymptotic value of the bag pressure ($B_{as}$) is analyzed w.r.t to the mass and radius of the HSs. The presence of hyperons in the hadronic sector also has significant effect on the choice of the value of $B_{as}$. Both the hadronic composition and the choice of $B_{as}$ significantly affect the stability of the star. The gross structural properties of the resultant HS are calculated in static condition and compared with the various constraints on them from different observational and empirical perspectives. The static properties like the maximum gravitational mass of the HS, obtained with $B_{as}$=80 MeV fm$^{-3}$, is consistent with the limits imposed from the observational analysis of PSR J0348+0432 and PSR J0740+6620. The estimates of $R_{1.4}$ and $R_{1.6}$ of HSs are found to be within the range prescribed from GW170817 analysis. Also the $M-R$ solutions of the HSs are in excellent agreement with the recently obtained NICER data for PSR J0030+0451. The results of maximum surface redshift, obtained with hybrid equation of state, satisfy the constraints from 1E 1207.4-5209 and RX J0720.4-3125. The work is also extended to obtain the tidal deformation properties of the HSs. The obtained value of $\Lambda_{1.4}$ is consistent with the bound obtained from GW170817 data analysis.

\end{abstract}

%

\section{Introduction}
\label{intro}

 Typical neutron star (NS) cores are characterized by extreme properties of matter like high density ($\sim 5-10$ times normal matter density $\rho_0 (=0.16$ fm$^{-3}$)) and very low temperature. At present heavy-ion collision experiments and lattice quantum chromodynamics (QCD) help in understanding the properties of hot and dense nuclear matter. NS cores are therefore the only possible environment suitable for the study of cold dense matter phases. Few recent astrophysical observations have not only remarkably added impetus to this direction but have also constrained the equation of state (EoS) of NS matter (NSM) to a considerable extent. The discovery of the massive pulsars like PSR J0348+0432 \cite{Ant} and PSR J0740+6620 \cite{Cromartie} have set strong upper bounds on the gravitational mass of NSs. Also, the ground breaking detection of gravitational waves (GW170817) from binary NS (BNS) merger by LIGO-Virgo collaboration, imposed stringent bounds on dimensionless tidal deformability ($\Lambda_{1.4}$) and radius ($R_{1.4}$) of a 1.4$M_{\odot}$ NS \cite{Abbott,Fattoyev,Most}. The detection of GW170817 also paved ways to constrain indirectly and co-relate several other NS properties like the symmetry energy \cite{sym_en}, bag constant with specific models \cite{Nandi,Nandi2,EnPingZhou,Rather}, speed of sound in NSM \cite{Kanakis-Pegios,NaZhang,cs3,Reed,Marczenko} and few others in terms of $\Lambda_{1.4}$ and $R_{1.4}$. Moreover, very recently Neutron star Interior Composition Explorer (NICER) experiment have come up with strong constraints on the $M-R$ relations of PSR J0030+0451 \cite{Riley,Miller}. Apart from the gravitational mass and radius, the maximum bounds on surface redshift ($Z_s$) are also obtained from the source spectrum analysis of 1E 1207.4-5209 \cite{Sanwal} and RX J0720.4-3125 \cite{Hambaryan}.

 Theoretical modeling of NSM at such densities are done to obtain realistic EoS that satisfy the aforesaid constraints on NS properties. Theoretically, such conditions are favorable for the formation of exotic matter like the hyperons \cite{Glendenning,Miyatsu2012,Bednarek2012,Weissenborn2012,Weissenborn2014,Agrawal2012, Lopes,Oertel,Colucci,Dalen,Lim,Rabhi2012,Sen2,Sen3}, delta baryons \cite{Glendenning,Cai,Sun,Kolomeitsev,Maslov,Zhu,Drago2014,Li2018,Sen,Sen4,Sen6}, various forms of boson condensates etc. \cite{Glendenning}. Deeper inside the core, the density becomes sufficiently high to even sustain deconfinement of hadronic matter to quark matter and consequent formation of hybrid stars (HSs) \cite{Glendenning,Weissenborn2011,Ozel2010,Klahn,Bonanno,Lastowiecki,Drago2016,Drago2016(2), Zdunik,Masuda,Wu,Sen,Sen2,Sen6,Most2,Zha}. The present work is dedicated to the study the HS properties in the light of the various constraints stated earlier. We adopt the effective chiral model \cite{Sahu2004,TKJ,Sen2,Sen4,Sen,Sen3,Sen5,Sen6} to describe pure hadronic phase while the MIT bag model \cite{Chodos,Glendenning} is considered to account for the pure quark phase. It is well known that the MIT bag model is characterized by the bag pressure $B$ which is actually the energy density difference between the perturbative vacuum and the true vacuum \cite{Burgio1,Burgio2}. Therefore it often taken as a free parameter whose value has a wide variation in literature. In the present work, the effective bag constant $B(\rho)$ is considered to have density dependence, following an Gaussian distribution as in \cite{Burgio1,Burgio2,Bordbar,Miyatsu2015}. As the density increases, a deconfinement transition from hadron to quarks is expected which will imply the vanishing of the difference between the perturbative and the non-perturbative (true) vacuum and hence the bag pressure $B$ should also vanish. This justifies strongly in favor of $B$ being a density dependent quantity, rather than being a constant. The dependence involves the values of $B(\rho)$ at asymptotic densities ($B_{as}$) and $B_0$ at $\rho=0$. In the present work, $B_{as}$ is varied within a limit that is consistent with the recent model dependent bounds specified from GW170817 \cite{Nandi,Nandi2,EnPingZhou,Rather}. The other constants in the parametrization has been adopted from \cite{Burgio2}.

 First order phase transition from hadronic to quark matter can be achieved in NS cores following Gibbs (GC) or Maxwell (MC) constructions. For a low value of surface tension $\sigma_s$ at hadron-quark transition interface, GC is favored with formation of stable mixed phase \cite{Glendenning,Orsaria2013,Rotondo,Sen,Sen2,Sen6} while for higher $\sigma_s$ ($\geq 70$ Mev fm$^{-3}$), MC is more relevant \cite{Maruyama,Maruyama2,Endo,Sotani,Shahrbaf,Xia2019} characterized by intermediate density jump between pure hadronic and pure quark phases \cite{Schramm,Lenzi,Bhattacharya,Logoteta2013,Sen,Sen2,Gomes2019,Han,Ferreira2020}. In the present work we adopt MC to achieve phase transition in dense HS core, assuming $\sigma_s$ to be sufficiently high.

 This paper is organized as follows. The hadronic model along with the model parameter set are discussed in nutshell with special emphasis on its main features in section \ref{Hadronic_model}. A touch of the attributes of the MIT bag model with density dependent bag pressure and the formalism for obtaining phase transition with MC is presented in section \ref{Quark phase}. A flavor of the mechanism to obtain the structural properties of HSs in static conditions and the tidal deformability is added in section \ref{stat_props}. We present the results and their detailed analysis in the following section \ref{Results}. The closing section \ref{conclusion} gives the summary of the present work.

\section{Formalism}
\subsection{Pure hadronic phase}
\label{Hadronic_model}

 For the hadronic phase, the effective chiral model \cite{Sahu2004,TKJ,Sen2,Sen4,Sen,Sen3,Sen5,Sen6} has been considered. Earlier in the same model, formation of the hyperons ($\Lambda,\Sigma^{-,0,+},\Xi^{-,0}$) has been taken into account along with the nucleons \cite{Sen2,Sen3}. This model based on chiral symmetry where the scalar $\sigma$ and pseudo-scalar $\pi$ mesons are the chiral partners. The dynamical generation of all the baryonic masses as well as that of the $\sigma$ and vector $\omega$ mesons is due to the spontaneous breaking of chiral symmetry at ground state \cite{Sen2,Sen4,Sen,Sen3,Sahu2004,TKJ,Sen5,Sen6}. Since in mean field approximation, $<\pi> = 0$ and the pion mass becomes $m_{\pi} = 0$, the explicit contributions are not there \cite{Sahu2004,TKJ,Sen2,Sen4,Sen,Sen3,Sen5,Sen6}. The isospin triplet $\rho$ mesons introduce the isospin asymmetry in the system. The Lagrangian contains an explicit mass term for the $\rho$ mesons following \cite{Sahu2004,TKJ,Sen2,Sen4,Sen,Sen3,Sen5,Sen6} unlike the dynamically generated masses of the $\sigma$ and $\omega$ mesons.

 $C_{\sigma N}$, $C_{\omega N}$, $C_{\rho N}$, $B$ and $C$ are the five model parameters. $B$ and $C$ are the coefficients of higher order scalar field terms while $g_{iN}$ are the meson-nucleon couplings, calculated in terms of $C_{iN}=g_i^2/m_i^2$, where $i=\sigma,\omega,\rho$. These five parameters are obtained by reproducing the symmetric nuclear matter (SNM) properties \cite{TKJ}. The same parameter set has been used in \cite{Sen,Sen2,Sen3,Sen4,Sen5,Sen6} to study dense matter properties relevant to NS and HS cores. It is adopted from \cite{TKJ} and tabulated below in table \ref{table-1} along with the SNM properties obtained with the particular parameter set.

\begin{table}[ht!]
\begin{center}
\caption{Parameter set of the nuclear matter model considered for the present work (adopted from \cite{TKJ}). The saturation properties such as binding energy per nucleon ($B/A$), nucleon effective mass ($m_N^{\star}$/$m_N$), the symmetry energy coefficient ($J$), slope parameter ($L_0$) and the nuclear matter incompressibility ($K$) defined at saturation density ($\rho_0$) are also listed.}

\setlength{\tabcolsep}{15.0pt}
\begin{center}
\begin{tabular}{cccccccc}
\hline
\hline
\multicolumn{1}{c}{$C_{\sigma N}$}&
\multicolumn{1}{c}{$C_{\omega N}$} &
\multicolumn{1}{c}{$C_{\rho N}$} &
\multicolumn{1}{c}{$B/m^2$} &
\multicolumn{1}{c}{$C/m^4$} &\\
\multicolumn{1}{c}{($\rm{fm^2}$)} &
\multicolumn{1}{c}{($\rm{fm^2}$)} &
\multicolumn{1}{c}{($\rm{fm^2}$)} &
\multicolumn{1}{c}{($\rm{fm^2}$)} &
\multicolumn{1}{c}{($\rm{fm^2}$)} & \\
\hline
6.772  &1.995  & 5.285 &-4.274   &0.292  \\
\hline
\hline
\multicolumn{1}{c}{$m_N^{\star}/m_N$}&
\multicolumn{1}{c}{$K$} & 
\multicolumn{1}{c}{$B/A$} &
\multicolumn{1}{c}{$J$} &
\multicolumn{1}{c}{$L_0$} &
\multicolumn{1}{c}{$\rho_0$} \\
\multicolumn{1}{c}{} &
\multicolumn{1}{c}{(MeV)} &
\multicolumn{1}{c}{(MeV)} &
\multicolumn{1}{c}{(MeV)} &
\multicolumn{1}{c}{(MeV)} &
\multicolumn{1}{c}{($\rm{fm^{-3}}$)} \\
\hline
0.85  &303  &-16.3   &32  &87  &0.153 \\
\hline
\hline
\end{tabular}
\end{center}
\protect\label{table-1}
\end{center}
\end{table}

 The nuclear incompressibility ($K=303$ MeV) is consistent with \cite{Stone,Stone2} but is little higher than that predicted by \cite{Garg}. An important feature of this model is that the effective mass of this model is dependent on both the scalar and vector fields. It is seen from \cite{Sahu2004,TKJ} that at high density, the dominance of vector potential increases. As a result, compared to other well-known RMF models, the nucleon effective mass ($m_N^{\star}=0.85~ m_N$) for this model is quite high and at high density unlike other RMF models, the effective mass increases after a certain value of density \cite{Sahu2004,TKJ,Sen2}. Moreover, the higher order terms of scalar field with coefficients $B$ and $C$ and the mass term of the vector field of the present model also become highly non-linear and dominant at high density. As a result the EoS softens at high density \cite{Sahu2004,TKJ,Sen2} and for the adopted parameter set the EoS passes through the soft band of heavy-ion collision data  for both SNM and pure neutron matter \cite{TKJ}. The EoS softens more when the formation of exotic baryons like hyperons and deltas are considered in NSM \cite{Sen,Sen2,Sen3,Sen4,Sen6}.

 The other SNM properties like the binding energy per nucleon ($B/A=-16.3$ MeV) and the symmetry energy ($J=32$ MeV), the saturation density ($\rho_0=0.153 \rm{fm}^{-3}$) and  are consistent with the estimates of \cite{Dutra2014,Khan2012,Khan2013}. The slope parameter ($L_0=87$ MeV) is a bit large compared to the findings of \cite{Tsang}. However, it is quite consistent with the range specified by \cite{Dutra2014}. Moreover, recent co-relation between the symmetry energy and tidal deformability and radius of a 1.4 $M_{\odot}$ NS came up to show that $L_0$ can be as high as $\sim 80$ MeV \cite{Fattoyev,Zhu2018}.

 This model and the adopted parameter set are thus well-tested and amply used to describe nuclear matter at finite temperature \cite{Sahu2004,Sen5} and dense matter compositions of NSs at zero temperature in the presence of $\Delta$ baryons, hyperons and quarks both in static and rotating cases \cite{Sen,Sen2,Sen3,Sen4,Sen6}. The overall detailed methodology to obtain the hadronic EoS with this model can be found in \cite{Sahu2004,TKJ,Sen,Sen2,Sen3,Sen4}. In the present work we have studied phase transition both in presence and absence of hyperons. Refs. \cite{Sahu2004,TKJ} describe the hadronic phase without hyperons while the mechanism to include hyperons in the same hadronic model can be found in \cite{Sen,Sen2,Sen3,Sen4,Sen6}.

 Similar to \cite{Weissenborn2012,Gupta,Glendenning,Glen2,Rufa,Sen2,Sen3,Sen4}, the hyperon-meson couplings $x_{iH}=g_{iH}/g_{iN}$ (where, $i=\sigma,\omega,\rho$ and $H=\Lambda,\Sigma,\Xi$) can be calculated in terms of the potential depths of the individual hyperon species \cite{Schaffner-Bielich,Sulaksono,Ishizuka,Sen2,Sen3,Sen4}). By fixing the value of scalar coupling constants $x_{\sigma_H}={g_{\sigma H}}/{g_{\sigma N}}$ consistent with the limit ($x_{\sigma_H} \leq 0.72$) specified by \cite{Glendenning,Glen2,Rufa}, we calculate the vector couplings $x_{\omega_H}={g_{\omega H}}/{g_{\omega N}}$ in terms of the binding energies of the individual hyperon species ($(B/A)_H|_{\rho_0}$ = -28 MeV for $\Lambda$, +30 MeV for $\Sigma$ and -18 MeV for $\Xi$ \cite{Schaffner-Bielich,Sulaksono}) in SNM, given by the following relation  \cite{Glendenning,Sen2,Sen3,Sen4}

\begin{eqnarray}         
(B/A)_H\biggr|_{\rho_0} = x_{\omega_H} ~ g_{\omega_N} ~\omega_0 + x_{\sigma_H}~ g_{\sigma_N} ~\sigma_0
\protect\label{be}
\end{eqnarray}         

 We have chosen $x_{\sigma_H}$=0.68 in the present work while $x_{\omega_H}$ is calculated by using the relation \ref{be}. $x_{\rho_H}$ is be chosen same as $x_{\omega_H}$ due to the close mass values of $\rho$ and $\omega$ mesons and also because both generate short range repulsive forces.

\subsection{Pure quark phase \& hadron-quark phase transition}
\label{Quark phase}

 The MIT bag model \cite{Chodos} with u, d and s quarks along with the electrons is considered to describe the pure quark phase. The u and d quark masses are much less compared to that of the s quark ($m_s \approx$ 95~MeV) \cite{Nakazato,PDG}. In its simplest form, it is considered that although the quarks are no longer confined within hadrons but they are constrained within a fully accessible `colorless' region known as the `Bag', characterized by a specific bag constant $B$ that determines the strength of quark interaction. This bag pressure is actually the energy density difference between the perturbative vacuum and the true vacuum \cite{Burgio1,Burgio2}. It is often taken as independent of density and the value of bag constant indeed plays a very decisive role in determining the properties of hybrid and quark stars. In the literature, \cite{Steiner,Buballa,Novikov,Baym} mentioned $B^{1/4}\sim ((100-300)$ MeV) while lattice calculations predict $B^{1/4}\sim 210$ MeV/fm$^3$ \cite{Benhar}. Recently, consistent with GW170817 observation, \cite{EnPingZhou} suggests that $B^{1/4} = (134.1-141.4)$ MeV for a low-spin prior while for the high-spin priors $B^{1/4} = (126.1-141.4)$ MeV considering pure quark stars with the MIT bag model. Ref. \cite{Nandi,Nandi2,Rather} also suggested similar maximum values of $B^{1/4}$ for HSs considering few well-known hadronic models for the hadronic phase. It is well established that higher values of bag constant gives stiffer EoS and results in more massive HSs \cite{Li2015,Logoteta2012,Tmurbagan,Yudin,Sen,Sen2}.

 As mentioned in the introduction section that in the present work we have considered density dependent bag pressure following the treatment of \cite{Burgio1,Burgio2}. The introduction of density dependence in the bag pressure takes into account the effects of increase in density as one approaches the core of the neutron/hybrid star. The density dependent bag pressure $B(\rho)$ attains finite values $B_0~\rm{at}~ \rho=0$ and $B_{as}$ at asymptotic densities. The density dependence of $B$ is given by a Gaussian distribution form in terms of $B_0$ and $B_{as}$ as \cite{Burgio1,Burgio2}

\begin{eqnarray}
B(\rho) = B_{as} + (B_0 - B_{as})~ \rm{exp}~ [-\beta(\rho/\rho_0)^2]
\label{B}
\end{eqnarray}

where, $\beta$ controls the decrease of $B$ with the increase of density. This  Gaussian form of the function driving the density dependence of the bag pressure involves the asymptotic behavior of the quarks at high densities relevant to HS cores. We intend to show that this can significantly affect the structural properties of HSs. Therefore we choose this specific form that involve the importance of $B_{as}$ at high density as given by \cite{Burgio1,Burgio2}. In the present work, we choose $B_0=400$ MeV fm$^{-3}$ and $\beta=0.17$ following \cite{Burgio1,Burgio2}. The precise value of $B_0$ is not at all important since there is no question of having any quark phase in this density domain. On the contrary, the asymptotic value of $B(\rho)$ is of greater significance with relevance to the EoS of the HSs. In order to invoke phase transition from hadronic to quark matter, we vary $B_{as}$ consistent with certain model dependent limits recently prescribed from GW170817 data analysis \cite{Nandi,Nandi2,EnPingZhou,Rather}. The effect of the change in $B_{as}$ values on the structural properties of the NSs has been studied.The main motivation of this work is to study the effects of this density dependence in contrast to the constant one. Moreover, from several analysis, it has been emphasized that the effects of perturbative corrections for the quark matter interactions on HS properties can also be realized by changing the bag constant \cite{Glendenning,Tmurbagan,Bhattacharya}. Hence we have not considered the contribution of the perturbative gluon exchange interactions separately in the present work.

 In the MIT Bag model, the energy density and pressure can be expressed as \cite{Glendenning}

\begin{eqnarray}
\varepsilon_{QM} = B(\rho)+ \sum_f \frac{3}{4\pi^2} \Biggl[\mu_fk_f(\mu_f^2-\frac{1}{2}m_f^2) - \frac{1}{2}m_f^4 \ln(\frac{\mu_f+k_f}{m_f})\Biggr]
 \protect\label{eos_e_uqm}
\end{eqnarray}

and

\begin{eqnarray}
P_{QM} = -B(\rho)+ \sum_f \frac{1}{4\pi^2} \Biggl[\mu_fk_f(\mu_f^2-\frac{5}{2}m_f^2)
+ \frac{3}{2}m_f^4 \ln(\frac{\mu_f+k_f}{m_f})\Biggr]
\protect\label{eos_P_uqm}
\end{eqnarray}

where,
\begin{eqnarray}
\mu_f=(k_f^2 + m_f^2)^{\frac{1}{2}}
\end{eqnarray}

and the total density is

\begin{eqnarray}
\rho=\sum_f \frac{k_f^3}{3\pi^2}
 \protect\label{density_upq}
\end{eqnarray}

 where, $f$ = u, d and s are the quark flavors. We impose the charge neutrality and $\beta$ equilibrium conditions on the system to obtain the number densities for each flavor.

 In the present work, first order phase transition is achieved from hadronic phase to quark phase using MC, assuming the surface tension at hadron-quark boundary to be sufficiently large \cite{Maruyama,Maruyama2}. The Maxwell criteria of local charge neutrality condition, which states that the individual phases must be charge neutral \cite{Schramm,Lenzi,Bhattacharya,Logoteta2013,Sen,Sen2,Gomes2019,Han, Ferreira2020}, is considered in order to achieve phase transition with MC. Phase transition with MC is achieved when the pressure and baryon chemical potential of the individual phases are equal. One therefore expects a jump in density when phase transition is invoked.

 We thus obtain the hybrid EoS and consequently proceed to study the structural properties of the HSs in static conditions.

\subsection{Neutron Star structure \& tidal deformability}
\label{stat_props}

 As the structural properties of NSs/HSs depend solely on the EoS, the former can be obtained once we compute the hybrid EoS.

 In static conditions the hybrid EoS is subjected to the Tolman-Oppenheimer-Volkoff (TOV) equations \cite{tov,tov2} that depict the hydrostatic equilibrium between gravity and the internal pressure of the star. Solving these equations, the static properties like central energy density ($\varepsilon_c$), gravitational mass ($M$), radius ($R$) and the surface redshift ($Z_s$) of the HS are calculated for the hybrid EoS.

 We also study tidal deformability of HSs. The tidal deformability parameter $\lambda$ is defined as the relation between the external tidal gravitational field ${\cal E}_{ij}$ exerted on the star and its deformation $Q_{ij}$. In the lowest order, the deformation $Q_{ij}$ may be considered quadrupolar, and

\begin{equation}
  Q_{ij} = -\lambda {\cal E}_{ij}.
\end{equation}

Then the value of $\lambda$ is related to the second (quadrupolar) tidal Love number $k_2$ as $\lambda = \frac{2}{3} k_2 R^5$. It is computed following the approach based on perturbation of the static metric in the Regge-Wheeler gauge of the star \cite{Hinderer2008,Hinderer2}. The resulting perturbation functions are solved along with the TOV equations to obtain the second Love number $k_2$. In the following, we will use the mass-normalized dimensionless value of the tidal deformability parameter

\begin{equation}
\Lambda = \frac{2}{3}k_2(R/M)^5.
\end{equation}

\section{Result and Discussions}
\label{Results}

 The hadronic EoS is obtained by following the methodology discussed in section \ref{Hadronic_model}. The hadronic matter is then allowed to undergo phase transition to quark matter in order to study the HS properties.

\subsection{Density dependence of bag pressure}

\begin{figure}[!ht]
\centering
\includegraphics[scale=1.0]{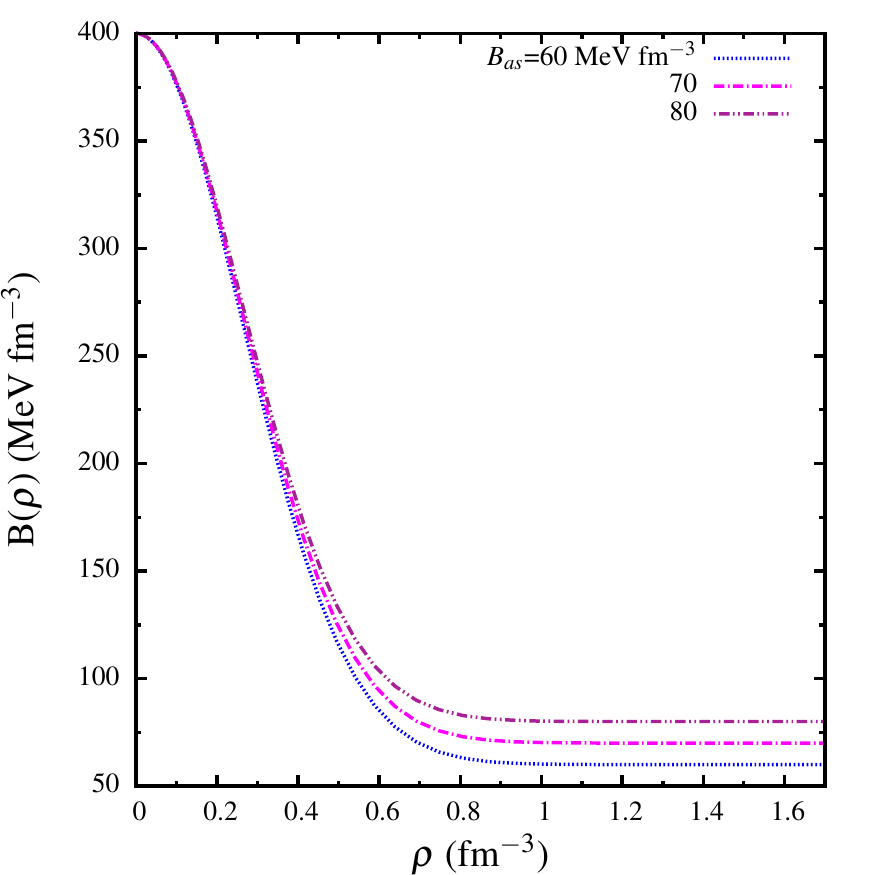}
\caption{\label{rB} Variation of bag pressure $B(\rho)$ with respect to density $\rho$ for different values of $B_{as}$.}
\end{figure}

 We invoked phase transition using the MIT bag model that involves density dependent bag pressure $B(\rho)$ following equation \ref{B} and the methodology discussed in section \ref{Quark phase}. In figure \ref{rB} we present the density dependence of the effective bag pressure $B(\rho)$ for different values of $B_{as}$. As mentioned in section \ref{Quark phase}, we choose $B(0)=400$ MeV fm$^{-3}$ following \cite{Burgio1,Burgio2}, the same is reflected in figure \ref{rB}. There is steep decrease of $B(\rho)$ upto certain values of baryon density $\rho$ depending on the chosen value of $\beta$, after which $B(\rho)$ shows no change with respect to $\rho$. The constant value that $B(\rho)$ attains thereafter denotes the chosen value of $B_{as}$ when quarks obtain asymptotic freedom.

\subsection{Hybrid stars with density dependent and density independent bag pressure}

\begin{figure}[!ht]
\centering
{\includegraphics[width=0.49\textwidth]{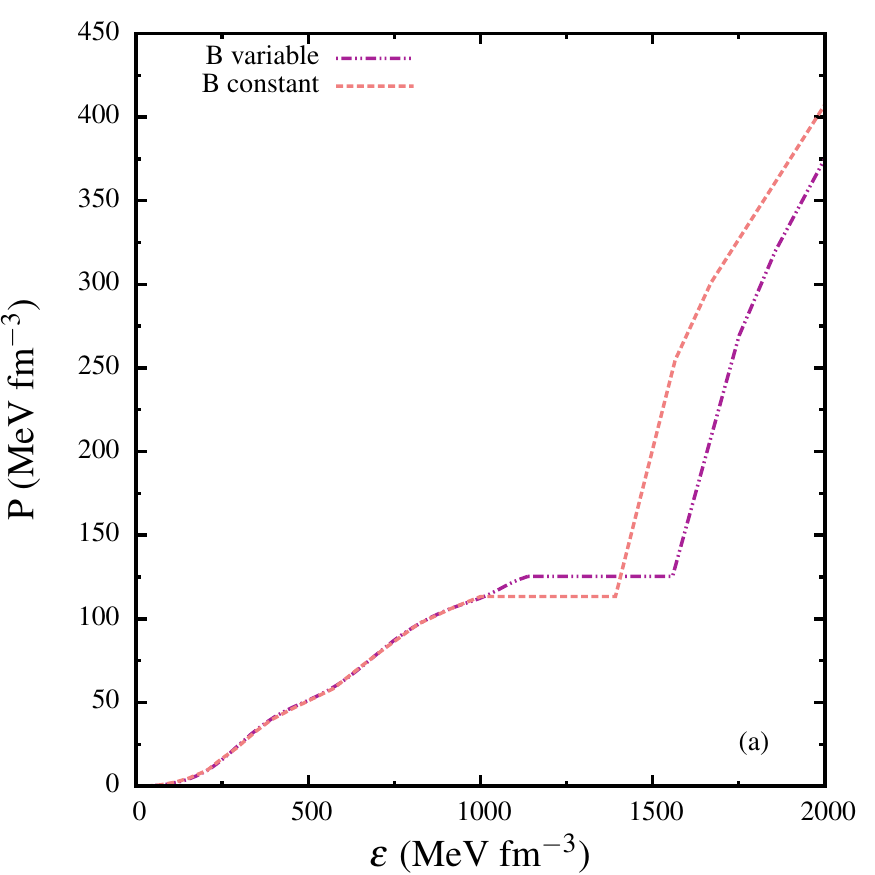}
\protect\label{eos1}}
\hfill
{\includegraphics[width=0.49\textwidth]{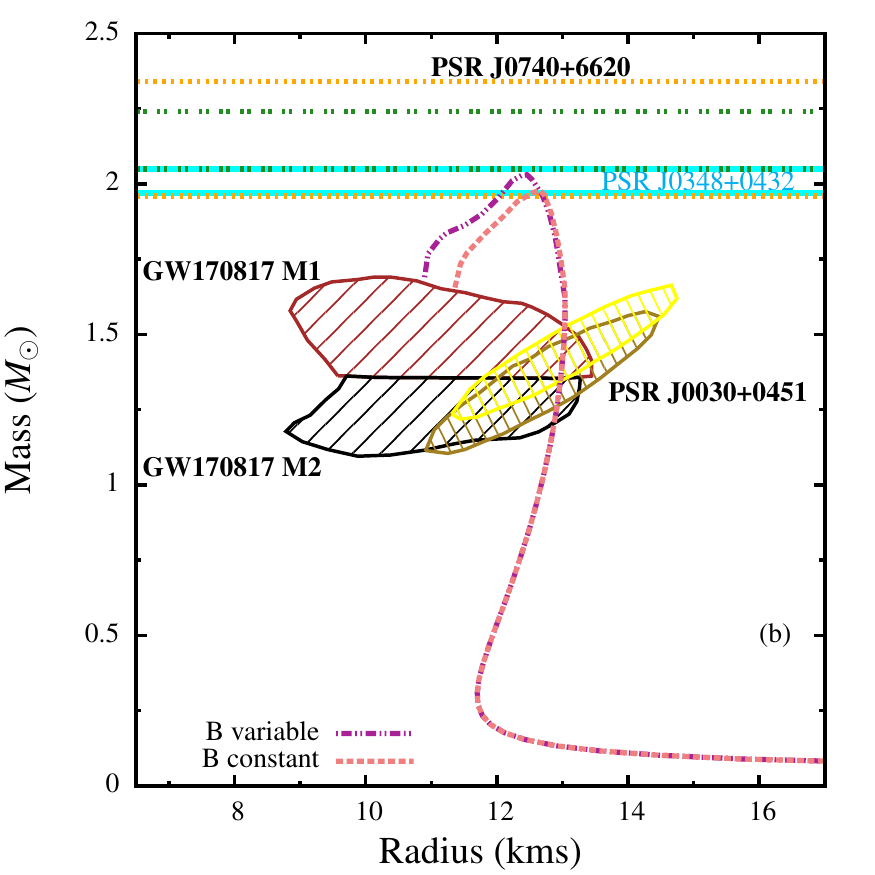}
\protect\label{mr1}}
\caption{\it (a) Equation of State of hybrid star matter for density dependent $B_{as}=80$ MeVfm$^{-3}$ and density independent bag pressure, $B=80$ MeVfm$^{-3}$ (left panel) and (b) the corresponding mass-radius
dependence of the hybrid stars (right panel). Observational limits imposed from high mass pulsars like PSR J0348+0432 ($M = 2.01 \pm 0.04~ M_{\odot}$) \cite{Ant} (area within cyan horizontal solid lines) and PSR J0740+6620 ($2.14^{+0.10}_{-0.09}~ M_{\odot}$ (68.3\% - area within dark-green horizontal double dotted lines) and $2.14^{+0.20}_{-0.18}~ M_{\odot}$ (95.4\% - area within orange horizontal single dotted lines)) \cite{Cromartie} are also indicated. The $M-R$ constraints (GW170817 M1 and GW170817 M2) from GW170817 \cite{Abbott} and from NICER experiment for PSR J0030+0451 \cite{Riley,Miller} are also compared.}
\protect\label{fig2}
\end{figure}

  With the pure quark matter EoS for the chosen values of $B_{as}$, we proceed to invoke phase transition. We first present the contrast between the two cases where the bag pressure is i) chosen constant ($B=$80 MeV fm$^{-3}$) and ii) density dependent with value of $B_{as}$=80 MeV fm$^{-3}$. The hybrid EoS is obtained for both the cases and compared in figure \ref{fig2} (left panel). The density dependent case is denoted by `$B$ variable' while the constant bag pressure case is denoted by `$B$ constant' in figure \ref{fig2}. As we adopted MC, a region of constant pressure is clearly seen for both the cases with corresponding jump in energy density. Such a result is consistent with works like \cite{Zdunik,Schramm,Zhang2016,Alvarez-Castillo2016,Blaschke2020,Ayriyan2019,Montana2019,Alvarez-Castillo2019,Paschalidis2018,Han,Ranea-Sandoval2019}. It is also evident from figure \ref{fig2} (left panel) that the transition is slightly delayed in case of density dependent bag pressure as compared to the case where the bag pressure is taken constant. The consequent effect is reflected in the HS properties like gravitational mass and radius shown in figure \ref{fig2} (right panel).

 The $M-R$ plot for the two cases clearly shows that the maximum gravitational mass of the HS with dependent bag pressure (2.03 $M_{\odot}$) is slightly more than that yielded by the density independent case (1.98 $M_{\odot}$) with corresponding radius being 12.67 km and 12.51 km, respectively. Similar result in terms of gravitational mass is also observed in \cite{Ritam}. In both the cases (B=constant and B=variable), the bounds on upper limit of maximum gravitational mass from PSR J0348+0432 \cite{Ant} and PSR J0740+6620 \cite{Cromartie} are satisfied. Moreover, the very recent $M-R$ constraints ($M=2.072^{+0.067}_{-0.066}~ M_{\odot}$ and $R=12.39^{+1.30}_{-0.98}$ km) obtained from modeling of NICER XTI event data for PSR J0740+6620 \cite{Riley2021} is well satisfied for `B=variable' case obtained with $B_{as}$=80 MeV fm$^{-3}$. The values of $R_{1.4}$ and $R_{1.6}$ in both cases are 13.00 km and 13.04 km, respectively and are well within the range prescribed for the same from GW170817 observation \cite{Abbott,Fattoyev,Bauswein}. Moreover, our estimates of gravitational mass and radius are well compatible with the recently obtained $M-R$ data for PSR J0030+0451 from NICER experiment \cite{Miller,Riley}.

\subsection{Hybrid star structure with density dependent bag pressure}

\begin{figure}[!ht]
\centering
{\includegraphics[width=0.49\textwidth]{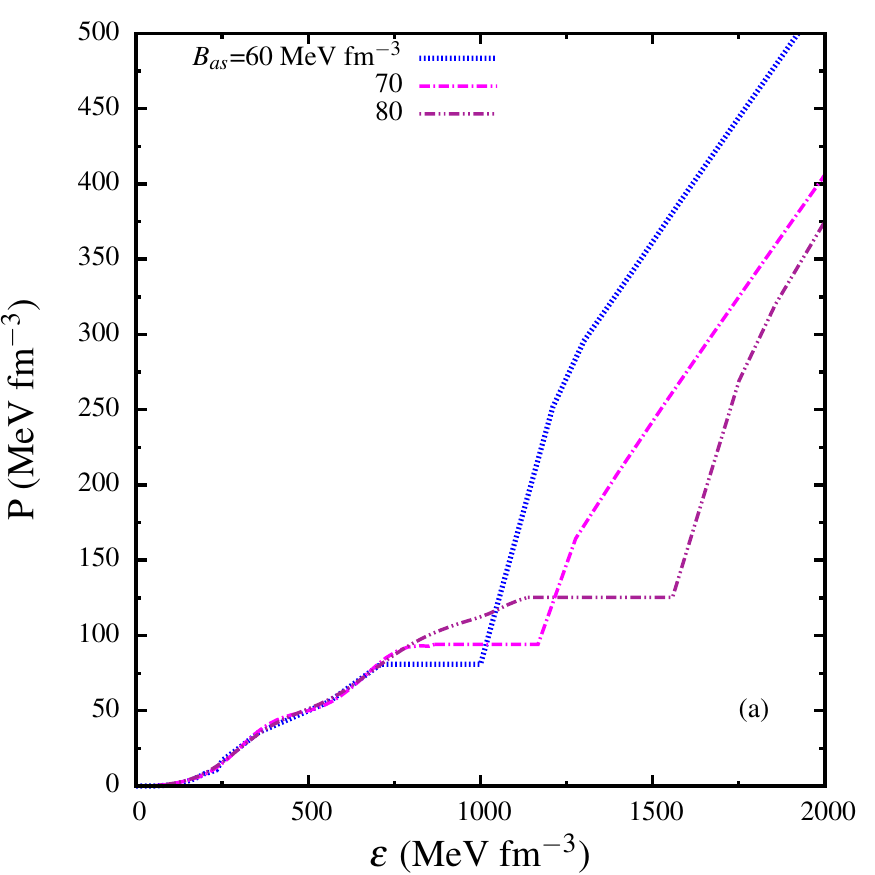}\protect\label{eos2}}
\hfill
{\includegraphics[width=0.49\textwidth]{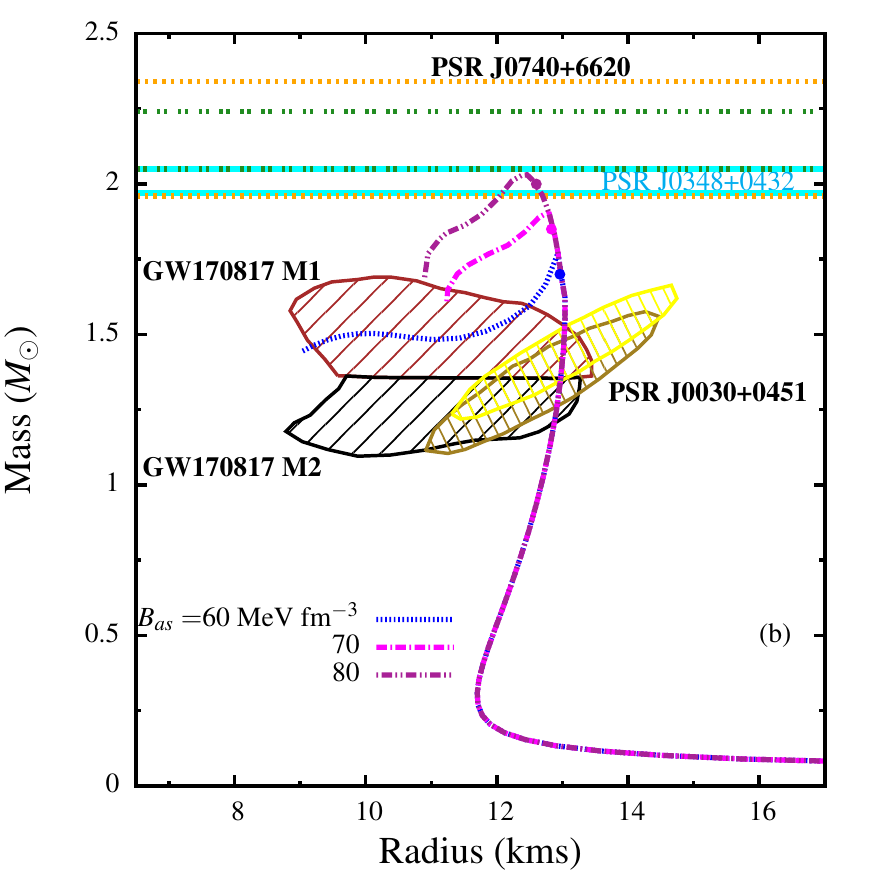}\protect\label{mr2}}
\caption{\it (a) Equation of State of hybrid star matter with density dependent bag pressure for different values of $B_{as}$ (left panel) and (b) the corresponding mass-radius dependence of the static hybrid stars (right panel). The various constraints on the $M-R$ plane are same as in right panel of figure \ref{fig2}. The marked points are corresponding mass of the hybrid stars where quarks start to nucleate.}
\protect\label{fig3}
\end{figure}

 We next proceed to study the HS properties for density dependent bag pressure with different values of $B_{as}$. We choose the value of $B_{as}$ as 60, 70 and 80 MeV fm$^{-3}$ in order to obtain the hybrid EoS (shown in left panel of figure \ref{fig3}). $B_0$ and $\beta$ still have their fixed values as 400 MeV fm$^{-3}$ and 0.17, respectively. As expected, the higher the value of bag pressure, the delayed and prolonged is the transition. For all values of $B_{as}$, the jump in energy density is due to MC and is clearly perceptible from \ref{fig3} (left panel). With these obtained hybrid EoS, we next investigate the static HS properties. The gravitational mass and radius yielded by the hybrid EoS are shown in right panel of figure \ref{fig3}.

 The variation in $B_{as}$ results in considerable change in the HS properties, specially the maximum gravitational mass. We obtain $M=1.77 M_{\odot}$ ($R=$13.00 km) for $B_{as}$=60 MeV fm$^{-3}$ and $M=1.90 M_{\odot}$ ($R=$12.83 km) for $B_{as}$=70 MeV fm$^{-3}$. For the last value of $B_{as}$(=80 MeV fm$^{-3}$), the maximum mass constraints from both PSR J0348+0432 \cite{Ant} and PSR J0740+6620 \cite{Cromartie} are satisfied. It also satisfies the very recent $M-R$ constraints obtained from NICER XTI event data for PSR J0740+6620 \cite{Riley2021}. Since phase transition occurs at high density, the HS solution with $B_{as}=$80 MeV fm$^{-3}$ attains maximum mass configuration soon after the onset of quarks. Therefore we achieve stable HSs within a small window of central density, beyond which the solution yields unstable HS whose mass drops down fast with decreasing values of radius.

  Another interesting feature is that we have obtained twin star configuration which is most prominent in case of the lowest value of $B_{as}$, clearly showing two maximas of gravitational mass (right panel of figure \ref{fig3}). A twin star configuration is usually achieved if the third stable sequence of compact stars is separated from the second one by an unstable region \cite{Hanauske2018,Alvarez-Castillo2017,Blaschke2016,Ayriyan2018,Alvarez-Castillo2019,Abgaryan,Maslov2019}. Twin stars thus belong to the third family of compact stars and there may be several types of them \cite{Christian,Montana2019,Christian2020,Blaschke2020,Alvarez-Castillo2021} depending on the conditions and the type of phase transition considered. Since phase transition occurs at higher densities, therefore for all the considered values of $B_{as}$, phase transition is reflected at high mass and therefore we obtain no change in the values of $R_{1.4}$ and $R_{1.6}$ with the variation of $B_{as}$. However, constraints on $R_{1.4}$ and $R_{1.6}$ from GW170817 observation \cite{Abbott,Fattoyev,Bauswein} are satisfied by all the HS configurations. Also the $M-R$ constraints on PSR J0030+0451 from both \cite{Miller,Riley} are satisfied with all the hybrid EoS.
  
\begin{figure}[!ht]
\centering
\includegraphics[scale=1.0]{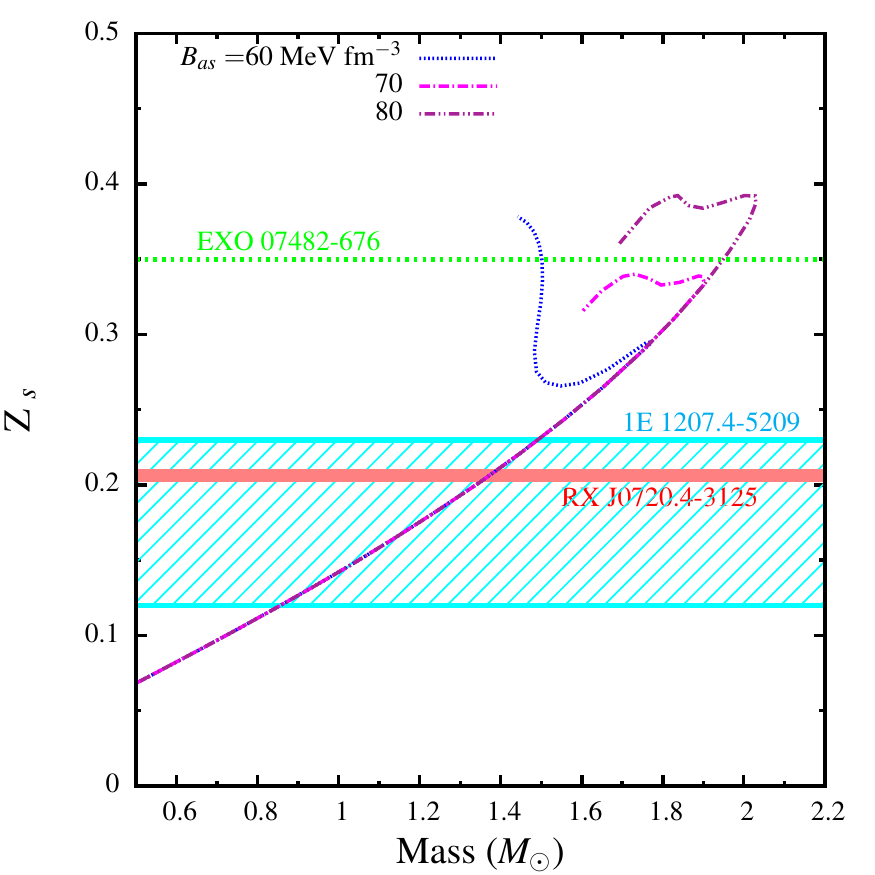}
\caption{\label{mZ} Surface gravitational redshift $Z_s$ vs gravitational mass $M$ of hybrid stars for different values of $B_{as}$. Observational limits imposed from EXO 07482-676, $Z_S = 0.35$ \cite{Cottam2002}, 1E 1207.4-5209, $Z_S = 0.12 - 0.23$ \cite{Sanwal} and RX J0720.4-3125, $Z_S = 0.205_{-0.003}^{+0.006}$ \cite{Hambaryan} are also indicated.}
\end{figure}

   We next present the variation of surface gravitational redshift $Z_s$ with respect to mass in figure \ref{mZ} yielded by the hybrid EoS with density dependent bag pressure as shown in the left panel of \ref{fig3}. From figure \ref{mZ} we find that the redshift is maximum (0.39) for the most massive HS configuration for $B_{as}$=80 MeV fm$^{-3}$. However, $Z_s$ for $B_{as}$=60 MeV fm$^{-3}$ is found to be more than that for $B_{as}$=70 MeV fm$^{-3}$ although the latter gives higher maximum mass than the former. This is because $Z_s$ depends both on the mass and radius. The hybrid EoS for $B_{as}$=60 MeV fm$^{-3}$ yields a twin star configuration and the lower maxima of the gravitational mass corresponds to a very small radius of 9.9 km. Therefore $Z_s$ for $B_{as}$=70 MeV fm$^{-3}$ is less than that of $B_{as}$=60 MeV fm$^{-3}$. For $B_{as}$=60 and 70 MeV fm$^{-3}$, the maximum values of $Z_s$ are 0.34 and 0.38, respectively. The values of maximum redshift obtained for all the hybrid EoS satisfy the observational bounds obtained from 1E 1207.4-5209 \cite{Sanwal} and RX J0720.4-3125 \cite{Hambaryan}. Although the estimates of $Z_s$ from EXO 07482-676 \cite{Cottam2002} are not confirmed, we have satisfied it for both $B_{as}$=60 and 80 MeV fm$^{-3}$.

\subsection{Tidal deformation of hybrid stars with density dependent bag pressure}

\begin{figure}[!ht]
\centering
{\includegraphics[width=0.49\textwidth]{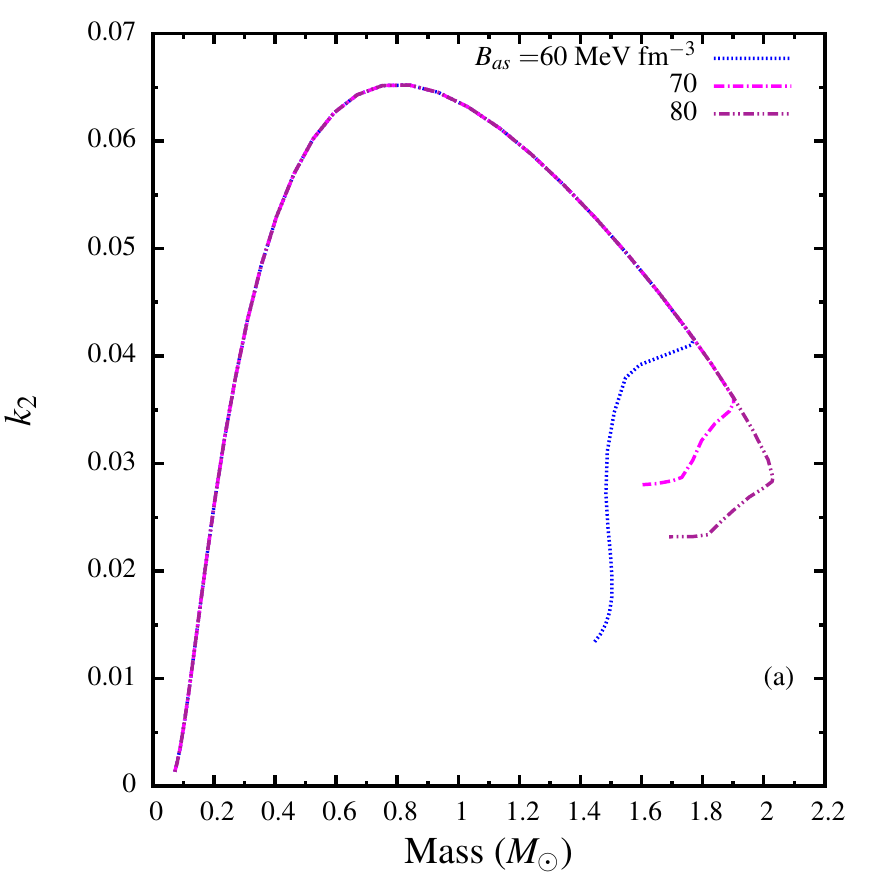}\protect\label{k2m}}
\hfill
{\includegraphics[width=0.49\textwidth]{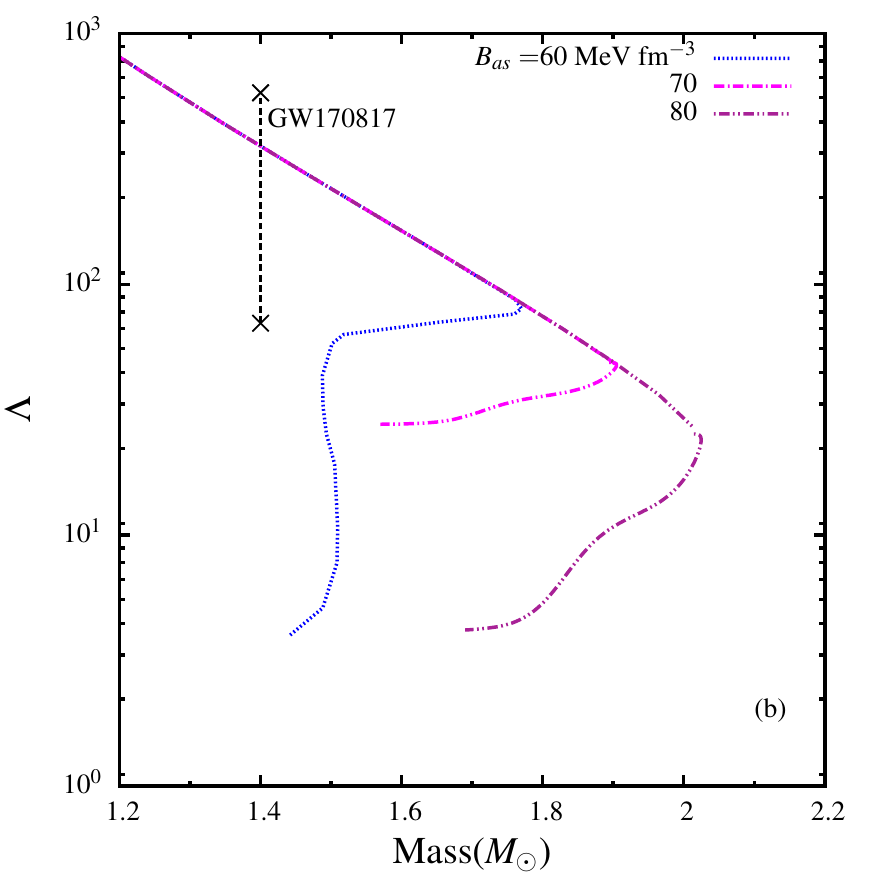}\protect\label{mLam}}
\caption{\it Variation of (a) second Love number (left panel) and (b) tidal deformability (right panel) with respect to gravitational mass of hybrid stars for different values $B_{as}$. Constraint on $\Lambda_{1.4}$ from GW170817 observations is also indicated following \cite{Abbott}.}
\protect\label{fig6}
\end{figure}

 The tidal deformation properties are studied next for the three HS configurations with density dependent bag pressure. The variations of second Love number $k_2$ and dimensionless tidal deformability $\Lambda$ with respect to the gravitational mass are shown respectively in the left and right panels of figure \ref{fig6}.

 The value of $k_2$ finds its maxima (0.065) at around $\sim 1 M_{\odot}$ indicating that for a given EoS, the quadrupole deformation is maximum for stars with intermediate mass values. The obtained values of $k_2$ are compatible with the range specified by \cite{Hinderer2008,Hinderer2}. As expected, $\Lambda$ shows considerable decrease for massive HSs. As mentioned earlier that we obtain phase transition at higher densities (higher mass) and therefore there is no change in the values of $k_2$ and $\Lambda$ corresponding to a 1.4 $M_{\odot}$ NS. For all values of $B_{as}$, we get $\Lambda_{1.4}=355.10$, which is well consistent with the constraints on $\Lambda_{1.4}$ obtained from GW170817 data \cite{Abbott}.

\begin{figure}[!ht]
\centering
\includegraphics[scale=0.55]{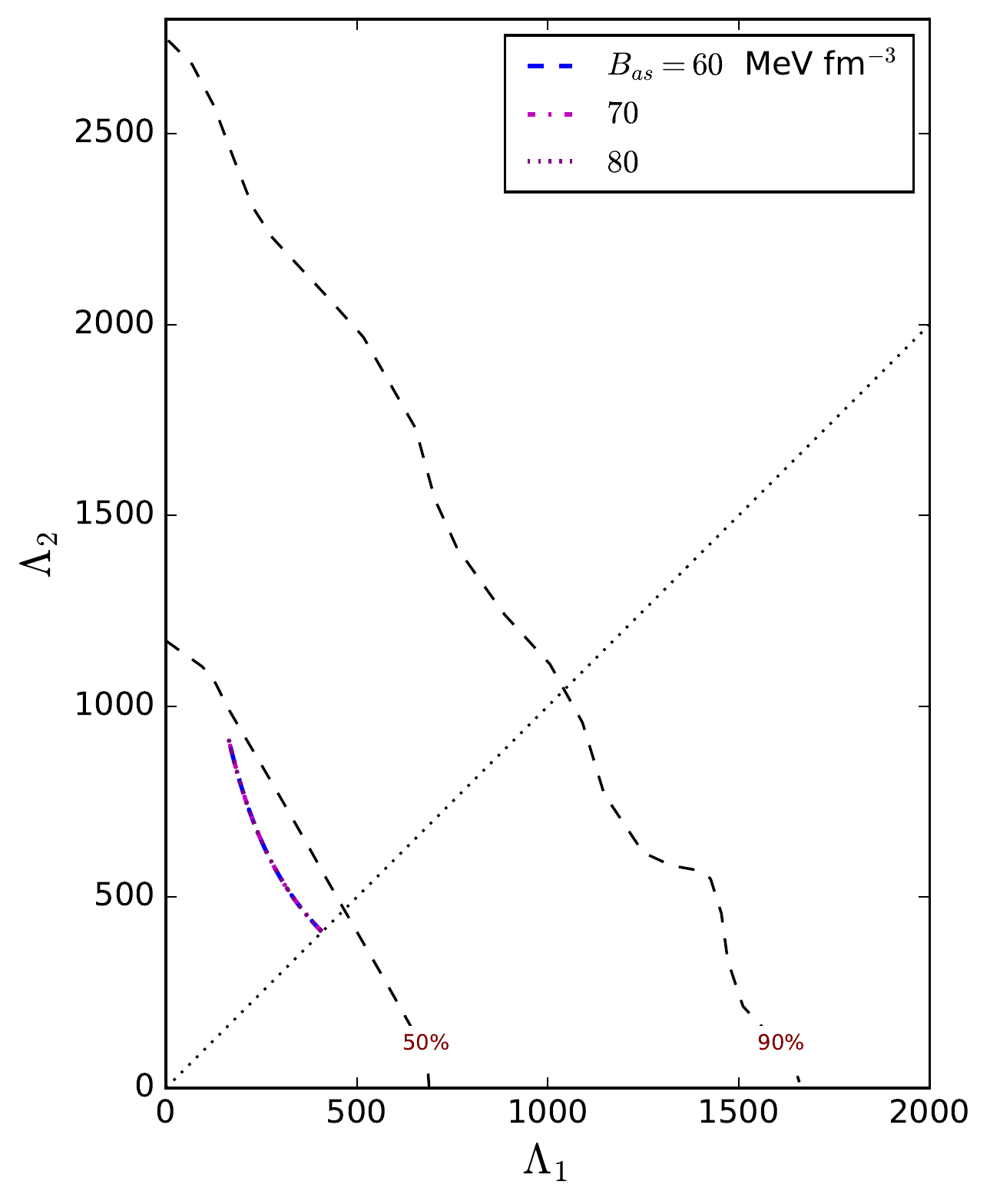}
\caption{\label{l1l2} Tidal deformabilities of the individual components of the BNS associated with GW170817 with hybrid equation of state for different values $B_{as}$. The 50\% and 90\% confidence limits for this event are also indicated following \cite{Abbott}.}
\end{figure}

 We finally show in figure \ref{l1l2} the the tidal deformability parameters $\Lambda_1$ and $\Lambda_2$ which are linked to the BNS companion having a high mass $M_1$ and a low mass $M_2$ associated with GW170817 observation. For the purpose the density dependent hybrid EoS for different values $B_{as}$ (as shown in left panel of figure \ref{fig3}) is employed.

 We find that the three curves for all the values of $B_{as}$ overlapping because as mentioned earlier that phase transition occurs at high density (high mass which is much more than the higher mass component of the binary 1.6 $M_{\odot}$ associated with GW170817). Our estimate of $\Lambda_1$ and $\Lambda_2$ is within the bounds specified from GW170817 data analysis \cite{Abbott}.

\subsection{Significance of $B_{as}$ and hadronic composition on stability of hybrid stars}

\begin{figure}[!ht]
\centering
{\includegraphics[width=0.49\textwidth]{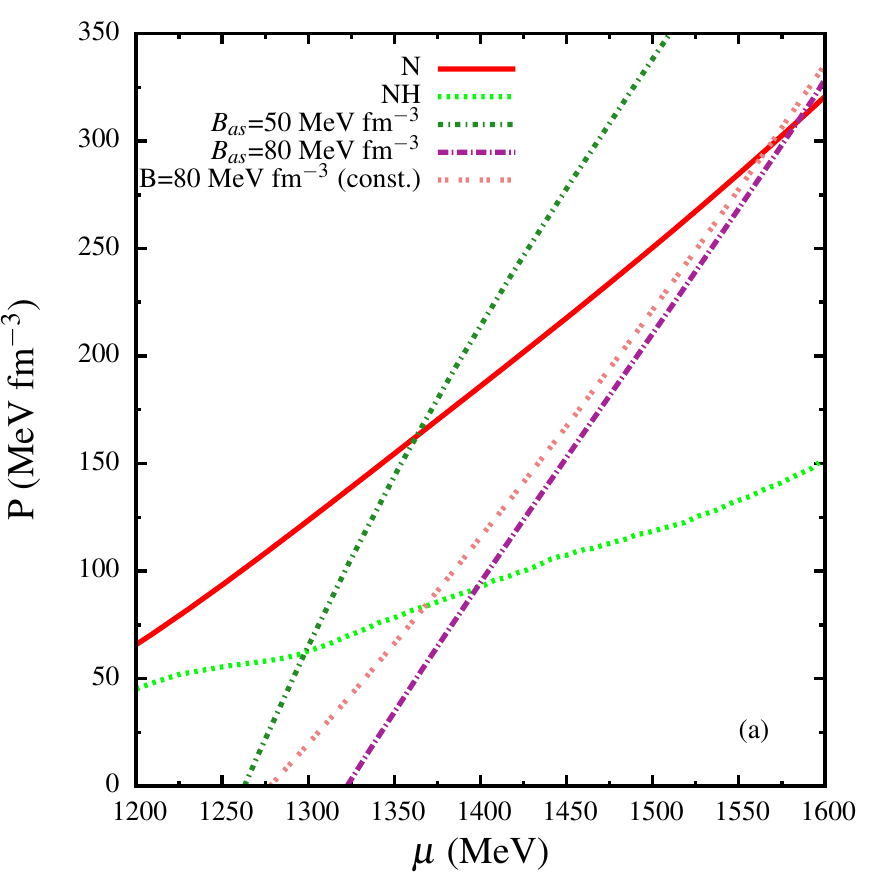}\protect\label{pressure}}
\hfill
{\includegraphics[width=0.49\textwidth]{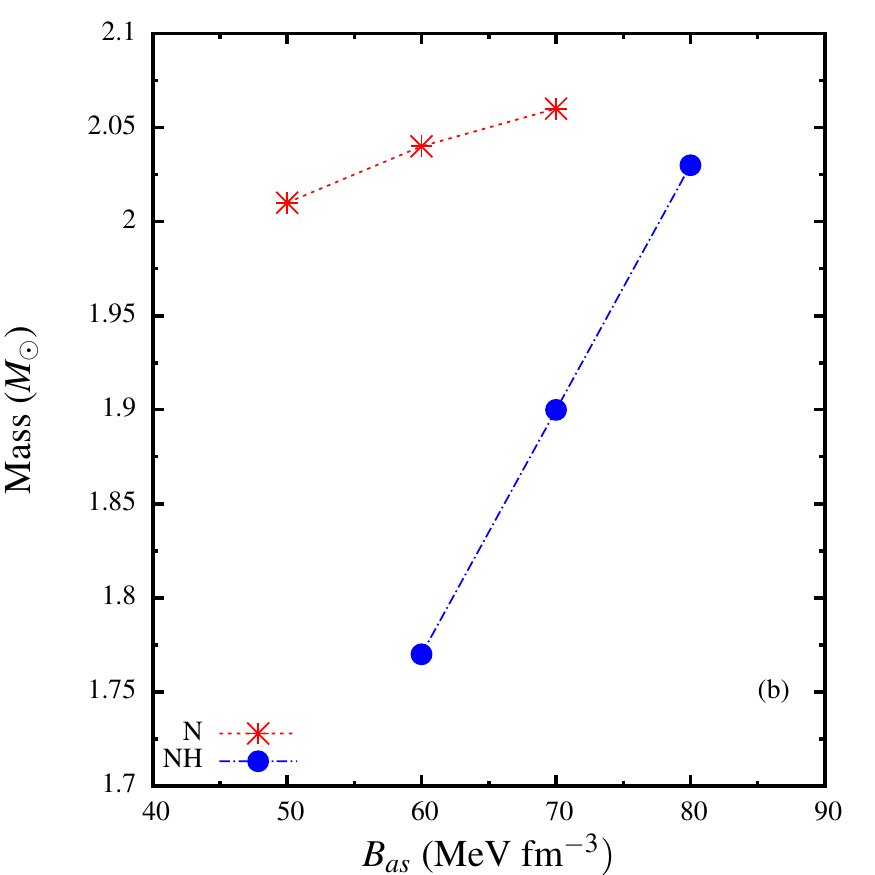}\protect\label{BasM}}
\caption{\it (a) Pressure vs chemical potential showing intersection of pure quark and pure hadronic equation of state with (NH) and without (N) hyperons (left panel). The quark equation of state for density independent bag pressure is also shown for comparison and (b) the variation of maximum gravitational mass $M$ with respect to $B_{as}$ of the hybrid stars with and without including hyperons (right panel).}
\protect\label{fig4}
\end{figure}

 In order to further illustrate the significance of the variation of $B_{as}$ on the hybrid EoS we have shown in figure \ref{fig4} (left panel) the pressure vs chemical potential ($\mu$) plot for different EoS. We have shown the cases of beta equilibrated NSM with nucleons and hyperons (NH) and also the one without including the hyperons (N). In the same plot \ref{fig4} (left panel), curves for the pure quark matter EoS with two values of $B_{as}$ as well as B=constant have been shown.

  The intersection points shown in \ref{fig4} (left) and in table \ref{table1} are the points in the $\mu-P$ plane where the hadronic and quark EoS intersect each other. The hadronic (${\rho_h}_{t}$) and quark densities (${\rho_q}_{t}$) at transition correspond to the value of chemical potential at transition ($\mu_t$). For higher values of $B_{as}$, the intersection is delayed in $\mu-P$ plane. 
  
   It is clear from the \ref{fig4} (left) that when the hyperons are not included (N), intersection takes place at higher chemical potential as compared to the case with hyperons (NH). For example, considering $B_{as}$=70 MeV fm$^{-3}$, the intersection occurs at $\mu_t=$1456.82 MeV for the case when hyperons are not included (N) while it is noticed at 1371.29 MeV in the presence of hyperons (NH) for the same value of $B_{as}$. This fact remains same for each value of $B_{as}$. Therefore comparatively delayed transition is expected for the case without hyperons. This will be manifested in the structural properties of the HSs. In order to demonstrate that, we have calculated the maximum mass corresponding to these cases and we have presented them in the right panel of figure \ref{fig4} and also in table \ref{table1}. In table \ref{table1} we have tabulated the intersection chemical potential ($\mu_t$), hadron (${\rho_h}_{t}$) and the quark (${\rho_q}_{t}$) densities at transition, mass of the HSs where quarks start to nucleate ($M_t$), maximum mass $M_{max}$ of HSs and corresponding central density ($\rho_c$) and the radius ($R$) for different values of $B_{as}$ with (NH) and without (N) hyperons.

\begin{table}[ht!]
\begin{center}
\caption{Hadron-quark intersection chemical potential ($\mu_t$), hadron (${\rho_h}_{t}$) and the quark (${\rho_q}_{t}$) densities at transition, mass of the hybrid stars where quarks start to nucleate ($M_t$), maximum mass $M_{max}$ of hybrid star and corresponding central density ($\rho_c$) and the radius ($R$) for different values of $B_{as}$ with (NH) and without (N) hyperons.}
\setlength{\tabcolsep}{7.0pt}
{\small{
\begin{center}
\begin{tabular}{ccccccccc}
\hline
\hline
\multicolumn{1}{c}{} &
\multicolumn{1}{c}{$B_{as}$} &
\multicolumn{1}{c}{$\mu_t$} &
\multicolumn{1}{c}{$({\rho_h}_{t})/\rho_{0}$} &
\multicolumn{1}{c}{$({\rho_q}_{t})/\rho_{0}$} &
\multicolumn{1}{c}{$M_t$} &
\multicolumn{1}{c}{$M_{max}$} & 
\multicolumn{1}{c}{$R$} &
\multicolumn{1}{c}{$\rho_c/\rho_{0}$} \\
\multicolumn{1}{c}{} &
\multicolumn{1}{c}{($MeV~fm^{-3}$)} &
\multicolumn{1}{c}{(MeV)} &
\multicolumn{1}{c}{} &
\multicolumn{1}{c}{} &
\multicolumn{1}{c}{($M_{\odot}$)} &
\multicolumn{1}{c}{($M_{\odot}$)} &
\multicolumn{1}{c}{(km)} &
\multicolumn{1}{c}{} \\
\hline
N  &50  &1362.50 &4.84 &6.44 &1.97 &2.01 &13.33 &6.49 \\
   &60  &1456.82 &5.62 &7.71 &2.02 &2.04 &13.19 &7.77 \\
   &70  &1528.03 &6.34 &8.85 &2.03 &2.06 &13.24 &8.87 \\
\hline
\hline
NH  &60           &1345.90 &4.64 &6.22 &1.73 &1.77 &13.00 &6.30 \\
    &70           &1364.84 &4.88 &6.48 &1.87 &1.90 &12.83 &6.53 \\
    &80           &1413.48 &5.38 &7.13 &2.00 &2.03 &12.67 &7.20 \\
    &80 (const.)  &1371.00 &4.97 &6.58 &1.94 &1.98 &12.51 &6.02 \\
\hline
\hline
\end{tabular}
\end{center}
}}
\protect\label{table1}
\end{center}
\end{table}

  As mentioned in the formalism section, we obtain phase transition following the Maxwell criteria where pressure and $\mu$ from each phase are equal and the hybrid EoS for the no-hyperon case (N) is also obtained by following the same procedure as that for the case where hyperons are included (NH). Therefore the hybrid EoS for the no-hyperon case (N) is also characterized by jump in energy density similar to the hybrid EoS obtained in presence of hyperons (NH) (as shown in left panels of figures \ref{fig2} and \ref{fig3}).
 
 With such hybrid EoS, we proceed to obtain the mass and radius estimates and the values of maximum mass obtained are shown in the right panel of figure \ref{fig4} and the table \ref{table1}. The delayed phase transition in case of HSs without hyeprons (N) results in higher values of the HS mass compared to that with hyperons (NH) (seen from the right panel of figure \ref{fig4} and the table \ref{table1}). For the hadronic EoS without the hyperons (N), maximum mass of $2.01 M_{\odot}$ is achieved for $B_{as}$ value as low as 50 MeV fm$^{-3}$. On the contrary, for the case including the hyperons (NH), this limit of $M\approx 2 M_{\odot}$ is reached at $B_{as} =80$. Also in this case with hyperons, the maximum mass ranges from 1.77 to 2.03 $M_{\odot}$ as $B_{as}$ varies from 60 to 80 MeV fm$^{-3}$ as seen from right panel of figure \ref{fig4} and table \ref{table1}. This variation in maximum mass is much less in the case that does not include the hyperons as is evident from the right panel of figure \ref{fig4}. Overall, for both the cases where hyperons are included (NH) or excluded (N) it is seen that for higher values of $B_{as}$, the hadronic phase continues much longer than the quark phase. This is also reflected in the $M-R$ solutions since in most cases the mass of the HSs where quarks start to nucleate ($M_t$) (marked in figure \ref{fig3} and mentioned in table \ref{table1}) and the maximum mass $M_{max}$ of HSs are found to be close. However, in all the cases $M_t$ is found to be lower than $M_{max}$ and the HS solutions in some cases are obtained within a small window of central density.
 
 In table \ref{table1} we report those values of $B_{as}$ that yield stable HS configurations. For no-hyperon case (N) we have shown the values of $B_{as}$ that yield HS configurations satisfying the maximum mass constraint ($M \geq 2 M_{\odot}$) from the most massive pulsars. For that case, choosing $B_{as}$ beyond 70 MeV fm$^{-3}$ no longer give stable stable HS solutions. Hence in case of ‘N’, we have not included further higher values of $B_{as}$ beyond 70 MeV fm$^{-3}$ in table \ref{table1}. However, it is seen that when hyperons are included (NH), stable HS solution is obtained upto $B_{as}$=80 MeV fm$^{-3}$. Although for $B_{as}$=80 MeV fm$^{-3}$, HS solution is obtained within a small window of central density, it yields the maximum mass of the HS consistent with the limit obtained from massive pulsar observations. We find that this is also true when bag parameter is independent of density i.e, $B$=80 (const.) MeV fm$^{-3}$ also yields stable HS configuration and the maximum mass satisfies the bound from massive pulsars (right panel of figure \ref{fig2}). However, considering the values of central density ($\rho_c$) of the individual stars presented in table \ref{table1}, it is seen that only with $B_{as}$=80 (const.) pure quark phase will not be formed at the center of the star since for this particular configuration $\rho_c$ lies between the hadronic (${\rho_h}_{t}$) and the quark (${\rho_q}_{t}$) densities at transition.
 
 Unlike the case where hyperons are included (NH) and for certain low values of $B_{as}$=60 MeV fm$^{-3}$, we do not obtain any prominent twin star configuration to account for third family of compact/hybrid stars in the case where hyperons are not included (N) for all the chosen values of $B_{as}$=50, 60 and 70 MeV fm$^{-3}$. This is due to the delayed transition in the latter case and the HS configurations become unstable shortly after stable maximum solutions are obtained. The solution becomes completely unstable if we further increase $B_{as}$ for the no-hyperon case (N) as the jump in energy density is too large in such cases. Moreover, large values $B_{as}$ leads to delayed phase transition as in the case of stiffer EoS without hyperons. The delayed phase transition leads to a large jump in energy density which is unfavorable for the formation of twin stars \cite{Christian2020,Khanmohamadi,Espino}.

 Since the transition takes place at higher densities in the no-hyperon case (N), therefore for a particular value of $B_{as}$, the maximum mass yielded in this case is comparatively more than the case when hyperons are included. For example at $B_{as}$=70 MeV fm$^{-3}$, the no-hyperon case gives $2.06 M_{\odot}$ while we obtain $1.90 M_{\odot}$ for the same value of $B_{as}$ when hyperons are included. It can be seen from the right panel of figure \ref{fig4} that maximum mass for no- hyperon (N) case is also comparatively less sensitive to the variation of $B_{as}$ as compared to the case when hyperons are included (NH). This is because of the delayed onset of quarks in the first case as a result of which the hybrid EoS stiffens more and the corresponding $M-R$ solution becomes unstable faster than that of the later case. Therefore we see that with increasing values of $B_{as}$ the HS configurations tend to become unstable for the no-hyperon case. At $B_{as}$=70 MeV fm$^{-3}$ the solution is almost unstable while at $B_{as}$=80 MeV fm$^{-3}$ it is completely unstable. However, in the case when hyperons are included, we still achieve stable HS for $B_{as}$=80 MeV fm$^{-3}$ though within a short range of central density (right panel of figure \ref{fig3}). Therefore the no-hyperon case yields massive HSs due to delayed phase transitions in their cores and soon become unstable with increasing values of $B_{as}$. This establishes that structural properties of HSs depend significantly on the values of $B_{as}$ and the composition of hadronic phase.

\section{Conclusion}
\label{conclusion}

 We study the possibility of hadron-quark phase transitions in HS cores.
The effective chiral model is adopted to account for the hadronic phase while the MIT bag model describes the quark phase. In the present work, density dependence of the bag pressure $B(\rho)$ is taken into account by varying $B_{as}$ which decides the density at which quarks attain asymptotic freedom. It is seen that the value of $B_{as}$ plays an important role in determining the HS properties especially the gravitational mass $M$. Compared to the density independent bag pressure case, we obtain 2.5\% increase in $M$ when density dependence is taken into account. It is also shown that the exclusion of hyperons from the hadron EoS can significantly lower the chosen maximum value of this parameter ($B_{as}$) while satisfying the upper limit constraints on maximum gravitational mass.

 In the present work MC is used to achieve phase transition and with $B_{as}$=80 MeV fm$^{-3}$ we have satisfied the constraints on maximum gravitational mass from PSR J0348+0432 and PSR J0740+6620. The chosen values of $B_{as}$ in the present work are consistent with certain model dependent bounds on the value of bag pressure prescribed from various recent works in the light of GW170817 data analysis.

 As phase transition is delayed in the present work, the radii estimates of $R_{1.4}$ and $R_{1.6}$ are found same for all the chosen values of $B_{as}$. Nevertheless the values $R_{1.4}$ and $R_{1.6}$ obtained for the HSs are consistent with the range suggested by analysis of GW170817 data from binary NS merger. Moreover, our $M-R$ solutions are in excellent agreement with the NICER data obtained for PSR J0030+0451. With the hybrid EoS, the constraints on maximum surface redshift $Z_s$ from 1E 1207.4-5209 and RX J0720.4-3125 are also satisfied.

  The tidal deformation properties of HSs are also studied in the present work. The value $\Lambda_{1.4}$, though found to be unaffected by our choice of $B_{as}$, lies well within the range specified by GW170817 data analysis. For the same reason we found the tidal deformabilities in the phase space of $\Lambda_1$ and $\Lambda_2$ (associated with the BNS of GW170817) overlapping in case of all the HS configurations achieved in this work with density dependent bag pressure.

 It is also seen that that structure and stability of HSs depend significantly on the values of $B_{as}$ and the composition of hadronic phase with respect to presence or absence of hyperons.

\ack

N. Alam and G. Chaudhuri acknowledge the support of "IFCPAR/CEFIPRA" project 5804-3.


\end{document}